\documentclass[english,aps,manuscript,aps,manuscript,showpacs]{revtex4}
\usepackage[T1]{fontenc}
\usepackage[latin9]{inputenc}
\usepackage{float}
\usepackage{amsthm}
\usepackage{amsmath}
\usepackage{graphicx}

\makeatletter
\@ifundefined{textcolor}{}
{%
 \definecolor{BLACK}{gray}{0}
 \definecolor{WHITE}{gray}{1}
 \definecolor{RED}{rgb}{1,0,0}
 \definecolor{GREEN}{rgb}{0,1,0}
 \definecolor{BLUE}{rgb}{0,0,1}
 \definecolor{CYAN}{cmyk}{1,0,0,0}
 \definecolor{MAGENTA}{cmyk}{0,1,0,0}
 \definecolor{YELLOW}{cmyk}{0,0,1,0}
}

\makeatother

\usepackage{babel}
\begin{document}

\title{Coherent response of a stochastic nonlinear oscillator to a driving
force: analytical characterization of the spectral signatures}

\author{J. Plata}

\address{Departamento de F\'{\i}sica, Universidad de La Laguna,\\
 La Laguna E38204, Tenerife, Spain.}
\begin{abstract}
We study the dynamics of a classical nonlinear oscillator subject
to noise and driven by a sinusoidal force. In particular, we give
an analytical identification of the mechanisms responsible for the
supernarrow peaks observed recently in the spectrum of a mechanical
realization of the system. Our approach, based on the application
of averaging techniques, simulates standard detection schemes used
in practice. The spectral peaks, detected in a range of parameters
corresponding to the existence of two attractors in the deterministic
system, are traced to characteristics already present in the linearized
stochastic equations. It is found that, for specific variations of
the parameters, the characteristic frequencies near the attractors
converge on the driving frequency, and, as a consequence, the widths
of the peaks in the spectrum are significantly reduced. The implications
of the study to the control of the observed coherent response of the
system are discussed. 
\end{abstract}

\pacs{05.40.Ca, 05.45. a, 89.75.Da}

\maketitle

\section{Introduction}

The combined effect of noise and deterministic driving can lead to
a variety of nontrivial features in the dynamics of nonlinear systems
\cite{key-bookMoss}. Remarkable effects emerging in different variations
of this scenario are stochastic resonance \cite{key-gammaitoniRMP},
noise-induced transport \cite{key-LandaPReport}, stochastic phase-switching
\cite{key-LapidusGabrielse,key-DykmanPRE,key-Brouardcollapses}, and
synchronization \cite{key-PikovskyBook,key-OjalvoPReport}. The interest
of the characterization of novel phenomenology in these systems is
clear: the combination of their fundamental components (i.e., nonlinearity,
driving forces, and fluctuations) can be relevant to widely different
contexts. (See \cite{key-CrossRoukes1,key-CrossRoukes2} for recent
developments in this field.) Here, we focus on an effect detected
in a recent realization of a basic model \cite{key-Stambaugh}. Namely,
we concentrate on the supernarrow peaks observed in the spectrum of
a nonlinear micro-mechanical torsional oscillator driven by a sinusoidal
force and subject to broadband noise of controllable intensity. Apart
from the standard $\delta$-component at the driving frequency, (associated
with the \emph{deterministic} component of the output), the spectral
density was found to display nontrivial supernarrow peaks emerging
from the (continuous) noisy background. Those peaks appeared at the
driving frequency, (for specific values of it), presented an asymmetric
form, and widened for increasing noise intensities. Secondary maxima
were also detected. The presence of those features was found to coincide
with the occurrence of approximately equal populations in the two
attractors existent  in the system at the considered frequency values.
(The term \emph{supernarrow peak} used along the paper is understood
as it is defined in \cite{key-DykmanPRL}: it refers to a peak whose
width is much smaller than the inverse of the relaxation time.) Here,
we intend to evaluate the connection of those effects with particular
aspects of the deterministic dynamics. In order to analytically establish
that link, we concentrate on a range of parameters where the nonlinear
term can be considered as a perturbation. In that regime, we present
a description of the dynamics based on averaging techniques applicable
to stochastic systems. From our approach, the mechanism responsible
for the detected behavior is uncovered and the elements of the system
which are essential for the appearance of the observed features are
identified. Furthermore, our semi-analytical picture of the dynamics
provides us with some clues to controlling the system response. The
direct implications of the study to the implementation of techniques
of stabilization of the system are evident. Additionally, because
of the fundamental character of the physics involved, the analysis
can be relevant to diverse  contexts, where conditions similar to
those realized in the basic scenario can be implemented. Here, it
is worth pointing out the existence of previous interesting work on
related systems \cite{key-DykmanPRL,key-DykmanKrivo}. In \cite{key-DykmanPRL},
an analytical approach to the spectral density was presented; in \cite{key-DykmanKrivo},
the activation energy between attractors in the configuration space
was evaluated. In our approach, a set of variables directly related
to those measured in the experiments is used. Moreover, we explicitly
apply the averaging methods that parallel the practical techniques.

The outline of the paper is as follows. In Sec. II, we present our
approach to the noisy dynamics of a damped and driven nonlinear oscillator.
Through the application of the averaging methods of Bogoliubov, Krylov,
and Stratonovich \cite{key-BogoliubovMethod,key-Stratonovich}, we
derive an effective description in terms of a system of stochastic
differential equations for the amplitude and the phase. In Sec. III,
the validity of our approach is confirmed through the qualitative
simulation of the main experimental findings. Moreover, we obtain
the spectrum for a linearized version of the model; from it, crucial
aspects of the role of noise in the response of the system to the
driving are identified. Finally, some general conclusions are summarized
in Sec. IV.

\section{The model system}

We consider a damped nonlinear oscillator driven by a sinusoidal force
and under the effect of additive noise. Specifically, we assume that
the evolution of the position coordinate $z$ (normalized to a typical
system length, and, therefore, dimensionless) is described by the
equation

\begin{equation}
\ddot{z}=-2\gamma\dot{z}-\omega_{0}^{2}z+\beta z^{3}+E\sin(\omega_{d}t)+\zeta(t),
\end{equation}
where the potential is characterized by the basic frequency $\omega_{0}$
and the nonlinear coefficient $\beta\,(>0)$. (Residual anharmonic
terms present in the experimental setup are not included as they were
shown to play no significant role in the detected features \cite{key-Stambaugh}.)
Additionally, $\gamma$ is the friction coefficient, $E$ and $\omega_{d}$
respectively stand for the amplitude and frequency of the driving
field, and $\zeta(t)$ denotes the stochastic force. In order to simulate
the experimental realization, $\zeta(t)$ is modeled as  general Gaussian
wideband noise \cite{key-RiskenFPequation}: the correlation function,
$k_{\zeta}(t^{\prime}-t)\equiv\left\langle \zeta(t)\zeta(t^{\prime})\right\rangle -\left\langle \zeta(t)\right\rangle ^{2}$,
is assumed to have a generic functional form, and the correlation
time is considered to be much shorter than any other relevant time
scale in the system evolution. The intensity coefficient $D=\frac{1}{2}\int_{-\infty}^{\infty}k_{\zeta}(\tau)d\tau$
will be used to characterize the noise strength \cite{key-Stratonovich}.
(The white-noise limit, defined by $k_{\zeta}(t^{\prime}-t)=2D\delta(t-t^{\prime})$,
is included in the analysis.) Additionally, a zero mean value, $\left\langle \zeta(t)\right\rangle =0$,
is considered. (A nonzero $\left\langle \zeta(t)\right\rangle $ can
be simply incorporated into the model as an effective deterministic
contribution.)

\subsection{The averaging method}

Our approach to deal with Eq. (1) is based on the coarse-graining
techniques developed by Krylov and Bogoliubov for the analysis of
deterministic nonlinear oscillations as they were adapted by Stratonovich
to the study of stochastic processes \cite{key-BogoliubovMethod,key-Stratonovich}.
Those averaging methods can be applied to generic wideband fluctuations
with sufficiently short correlation time. In this approach, the amplitude
$A$ and the phase $\Psi$ of the oscillations are defined through
the equations

\begin{eqnarray}
z & = & A\cos(\omega_{d}t+\Psi)\nonumber \\
\dot{z} & = & -\omega_{d}A\sin(\omega_{d}t+\Psi)
\end{eqnarray}
With these changes, Eq. (1) is reduced to a system of two first-order
equations in \textit{standard form} \cite{key-Stratonovich,key-plataPRElanda},
i.e., with a structure corresponding to a harmonic oscillator perturbed
by deterministic and stochastic terms. For $\omega_{d}\gg\gamma$,
the average of the deterministic perturbative elements over the period
$\tau_{d}=2\pi/\omega_{d}$ is readily carried out. Moreover, for
a noise correlation time much smaller than the relaxation times of
the amplitude and the phase, the coarse graining of the stochastic
terms over $\tau_{d}$ can be applied following the procedure presented
in \cite{key-Stratonovich}. Accordingly, we obtain that, to first
order, the averaged equations are \cite{key-Stratonovich,key-plataPRElanda}

\begin{equation}
\dot{A}=-\gamma A-\frac{1}{2}\frac{E}{\omega_{d}}\cos\Psi+\frac{1}{4}\frac{D_{eff}}{\omega_{d}}\frac{1}{A}+\xi_{1}(t),
\end{equation}

\noindent 
\begin{equation}
\dot{\Psi}=\Delta-\frac{3}{8}\frac{\beta}{\omega_{d}}A^{2}+\frac{1}{2}\frac{E}{\omega_{d}}\frac{1}{A}\sin\Psi+\frac{\xi_{2}(t)}{A},
\end{equation}

\noindent where we have incorporated the detuning $\Delta=\omega_{0}-\omega_{d}$
and have assumed that $\omega_{0}+\omega_{d}\simeq2\omega_{d}$. $\xi_{1}(t)$
and $\xi_{2}(t)$ are effective Gaussian white-noise terms defined
by $\left\langle \xi_{i}(t)\right\rangle =0$ and $\left\langle \xi_{i}(t)\xi_{j}(t^{\prime})\right\rangle =2D_{eff}\delta_{i,j}\delta(t-t^{\prime})$,
($i,j=1,2$), with $D_{eff}=\kappa_{\zeta}(\omega_{d})/(4\omega{}^{2})$.
{[}$D_{eff}$, which determines the strength of the (uncorrelated)
effective noise terms, is obtained from the power spectral density
$\kappa_{\zeta}(\omega)\equiv\int_{-\infty}^{\infty}e^{i\omega\tau}k_{\zeta}(\tau)d\tau$
of the original noise $\zeta(t)$ at the frequency $\omega_{d}$.
From the broadband characteristics assumed for $\zeta(t)$, a smooth
form of $\kappa_{\zeta}(\omega)$ can be inferred. Indeed, a completely
flat spectrum occurs in the white-noise limit.{]} Whereas the noise
term in Eq. (3), $\xi_{1}(t)$, is additive, the fluctuations enter
Eq. (4) through the term $\xi_{2}(t)/A$, and, therefore, have multiplicative
character. Moreover, it is important to take into account the presence
of the noise-induced \emph{deterministic} term $\frac{1}{4}\frac{D_{eff}}{\omega_{d}}\frac{1}{A}$
in Eq. (3). Because of this term, the point defined by $A=0$ is never
reached in the stochastic dynamics. (We stress that the emergence
of this term is due to the multiplicative character of noise in the
set of variables corresponding to the amplitude and phase \cite{key-Stratonovich}.)
The use of averaged equations is specially appropriate for the considered
experimental setup: the specific characteristics of the applied detection
scheme imply that the registered data do actually correspond to averaged
magnitudes. (Equations with a similar form have been extensively studied
by Stratonovich in the context of nonlinear self-excited oscillations
in electronic circuits \cite{key-Stratonovich}.)

\subsection{The deterministic dynamics}

In order to trace the response of the system to noise, we must clearly
define the deterministic scenario into which the fluctuations enter.
(See \cite{key-CrossBook} for an exhaustive presentation of standard
techniques applicable to this context.) The noiseless dynamics is
described by Eq. (1) without the random term $\zeta(t)$, and, consequently,
by Eqs. (3) and (4) with $D_{eff}=0$. The stationary values of the
amplitude $A_{s}$ and phase $\Psi_{s}$ are obtained by setting $\dot{A}=0$,
and $\dot{\Psi}=0$ in Eqs. (3) and (4). After minor algebra, it is
found that $A_{s}$ is obtained from the roots $x_{s}$ of the equation 

\begin{equation}
(1-x)^{2}x=c_{1}-c_{2}x
\end{equation}
where we have introduced the coefficients $c_{1}=\frac{3}{32}\frac{E^{2}\beta}{\omega_{d}^{3}\Delta^{3}},$
and $c_{2}=-\frac{\gamma^{2}}{\Delta^{2}}$, and have scaled the squared
amplitude as $x=\frac{3}{8}\frac{\beta}{\omega_{d}\Delta}A^{2}$.
Additionally, the stationary phase $\Psi_{s}$ is given by 

\begin{equation}
\tan\Psi_{s}=\frac{\Delta}{\gamma}(1-x_{s})
\end{equation}
The graphical resolution of Eq. (5) is illustrated in Fig. 1 for different
values of the driving frequency. The fixed points correspond to the
intersections between the graphics of the functions $f(x)\equiv(1-x)^{2}x$
and $g(x)\equiv c_{1}-c_{2}x$. Their number, localization, and stability
are strongly dependent on $\omega_{d}$, which enters the coefficients
$c_{1}$, $c_{2}$, and the scaling factor $\frac{x}{A^{2}}=\frac{3}{8}\frac{\beta}{\omega_{d}\Delta}$,
directly, and/or, through the detuning. Let us look at the stability.
The analysis of the linearized equations is straightforward: taking
$A=A_{s}+\delta A$, and $\Psi=\Psi_{s}+\delta\Psi$, and defining
$u\equiv\delta A$ and $v\equiv A_{s}\delta\Psi$, we find 

\begin{eqnarray}
\dot{u} & = & -\eta_{1}u+\lambda_{1}v,\\
\dot{v} & = & \lambda_{2}u-\eta_{2}v.
\end{eqnarray}

\noindent where the diagonal elements read $\eta_{1}=\eta_{2}=\gamma$,
and the coupling terms are 

\begin{eqnarray}
\lambda_{1} & = & \Delta(x_{s}-1)\nonumber \\
\lambda_{2} & = & \Delta(1-3x_{s}).
\end{eqnarray}

\noindent The characteristic frequencies $\Omega_{\pm}$ are trivially
given by the expression 

\begin{eqnarray}
\Omega_{\pm} & = & \frac{1}{2}\left[-(\eta_{1}+\eta_{2})\pm\sqrt{(\eta_{1}-\eta_{2})^{2}+4\lambda_{1}\lambda_{2}}\right],
\end{eqnarray}
which for the considered parameters becomes

\begin{equation}
\Omega_{\pm}=-\gamma\pm\Delta\sqrt{(x_{s}-1)(1-3x_{s})}.
\end{equation}

\noindent Then, the time evolution is given by $u(t)=G_{+}e^{\Omega_{+}t}+G_{-}e^{\Omega_{-}t}$,
and by a similar equation for $v(t)$. ($G_{+}$ and $G_{-}$ are
constants determined by the initial conditions.) Moreover, for the
original system, connected with its coarse-grained reduction through
the change of variables introduced in Eq. (2), we can write

\begin{eqnarray}
z(t) & = & (A_{s}+u)\cos(\omega_{d}t+\Psi_{s}+\frac{v}{A_{s}})\nonumber \\
 & \approx & A_{s}\cos(\omega_{d}t+\Psi_{s})+u\cos(\omega_{d}t+\Psi_{s})-v\sin(\omega_{d}t+\Psi_{s}).
\end{eqnarray}
Therefore, apart from a constant-amplitude oscillation with frequency
$\omega_{d}$, $z(t)$ incorporates two terms with time-dependent
amplitudes $u(t)$ and $v(t)$. In a generic regime, since the frequencies
$\Omega_{\pm}$ can be complex, the global output signal can have
contributions centered at frequencies shifted from $\omega_{d}$,
coming from the modulation terms. In the particular regimes where
the frequencies $\Omega_{\pm}$ take negative real values, the modulated
oscillations are merely damped signals at the external frequency.
The precise identification of the different regimes will be crucial
for explaining the coherent response of the stochastic system to the
driving.

From Eq. (10), it is apparent that necessary and sufficient conditions
for stability, i.e., for both frequencies to have negative real parts,
(and, consequently, for the solutions to decay in time), are

\begin{equation}
\eta_{1}+\eta_{2}>0,
\end{equation}
and

\begin{equation}
\eta_{1}\eta_{2}-\lambda_{1}\lambda_{2}>0.
\end{equation}
For the parameters of our system, Eq. (13) trivially holds and Eq.
(14) is cast into 
\begin{equation}
\gamma^{2}-\Delta^{2}(x_{s}-1)(1-3x_{s})>0.
\end{equation}
As the driving frequency is varied, $\Delta$ and $x_{s}$ are modified,
and, in turn, the stability of the system is affected. This is illustrated
in Figure 1, where we depict a sequence of representative solutions
obtained as $\omega_{d}$ is increased. \bigskip{}

\includegraphics[scale=0.6]{Figure1}

\begin{figure}[H]
\caption{Graphical resolution of Eq. (5) for different values of the driving
frequency: $\omega_{d}=0.902\; Hz$ (a); $\omega_{d}=0.9056\; Hz$
(b); $\omega_{d}=0.9065\; Hz$ (c); $\omega_{d}=0.909\; Hz$ (d).
(The rest of parameters are: $\omega_{0}=1.\; Hz$, $\beta=3.1\times10^{-2}\; s^{-2}$,
$\gamma=5.1\times10^{-2}\; s^{-1}$, and $E=2.4\times10^{-1}\; s^{-2}$.
This set of parameters is used in all the Figures.) }
\end{figure}

The system can present monostable or bistable character depending
on the combinations of parameters incorporated into the coefficients
$c_{1}$, $c_{2}$. (Following the experimental scheme, in the considered
cases, only the driving frequency is varied; the rest of parameters
are fixed.) Fig. 1a represents a situation where there is only one
attractor, i.e., one stable solution, (the point $P$). As the frequency
is increased, a setting similar to that considered in the experiments
emerges. Namely, as shown in Fig. 1b, the system presents then three
fixed points: two attractors, (the points $P$ and $Q$), and one
unstable solution, (the point $R$.) The corresponding basins of attraction
are depicted in Fig. 2. {[}An effective phase space ($q,p$) has been
standardly defined for the coarse-grained scenario by taking $q=A\cos\Psi$,
and $p=A\sin\Psi$.{]} Depending on the initial conditions, the system
eventually reaches one or other stable point. For the considered set
of parameters, the basins have similar sizes. Hence, a uniform initial
distribution of population in the phase space eventually evolves into
an approximately symmetric bistable distribution in the attracting
sites. Fig. 1c, shows that $P$ comes near to $R$ as the external
frequency is increased. As observed in Fig. 1d, further growth of
$\omega_{d}$ induces the disappearance of $P$ after converging on
the unstable point. (It must be taken into account that the limit
of the stability region is reached when the equation $\Delta^{2}(x_{s}-1)(1-3x_{s})-\gamma^{2}=0$
is fulfilled.) \bigskip{}

\includegraphics[scale=0.5]{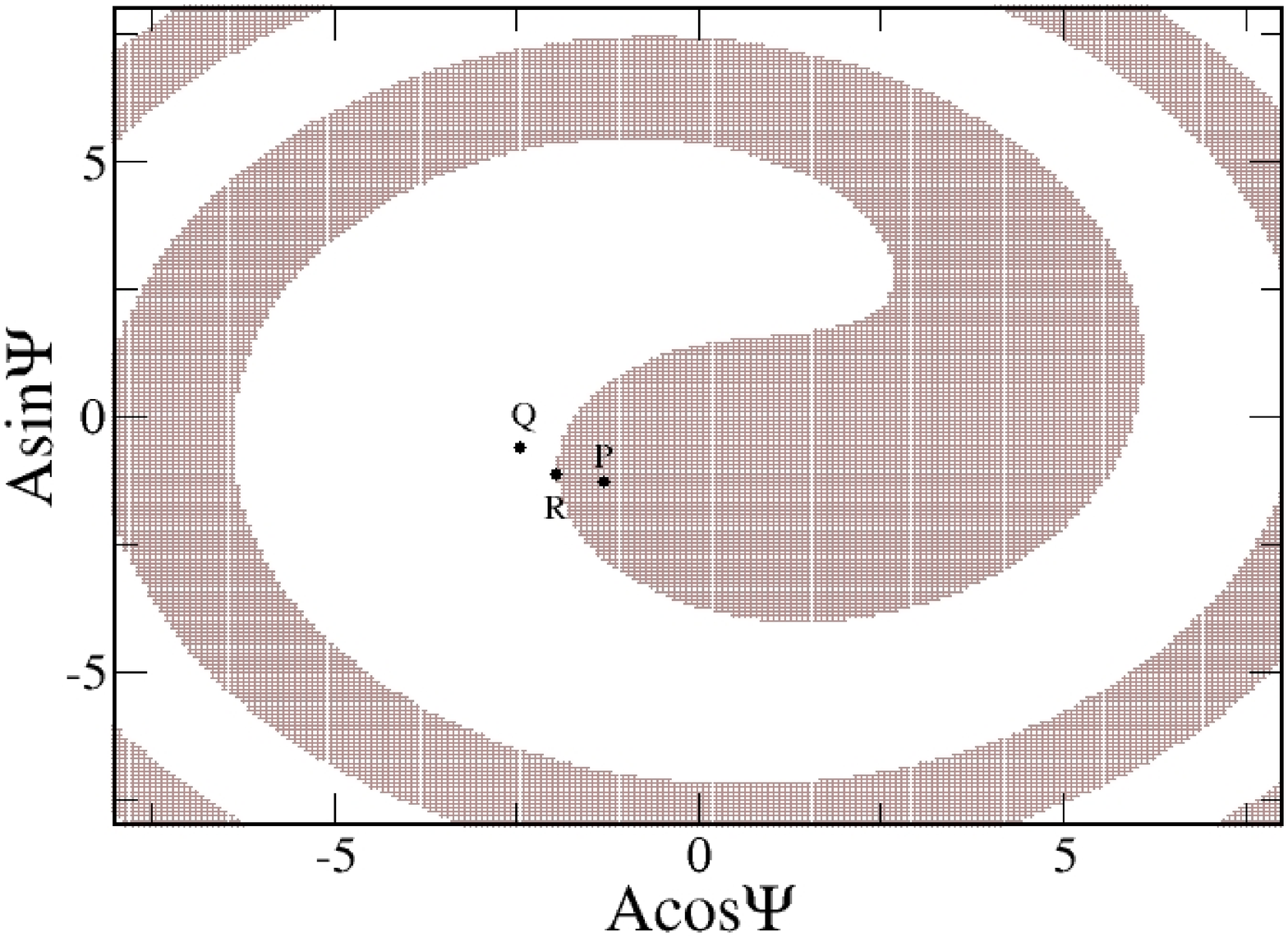}

\begin{figure}[H]
\caption{Effective phase space of the averaged system. $P$ and $Q$ denote
(stable) attractors, and $R$ represents an unstable solution. The
set of parameters is the same as in Fig. 1b. (Arbitrary units are
used for the amplitude.)}
\end{figure}

Closely related to the above analysis is a nontrivial characteristic
of the original system. Namely, as reported in \cite{key-Stambaugh},
the system presents hysteresis in the \emph{evolution} of the stationary
amplitudes as the external frequency is varied. Specifically, the
stationary values of the amplitude corresponding to an adiabatic sequence
of increasing driving frequencies differ from those corresponding
to the inverse (frequency-decreasing) sequence. Fig. 3 shows that
this aspect of the dynamics is satisfactorily reproduced in the averaged
scenario. There, the arrows indicate the \emph{paths} followed by
the amplitude as the frequency is increased (or decreased.) The \emph{jumps}
in amplitude correspond to the appearance (or disappearance) of the
stable fixed points. (Note that as we are focusing on the jumps, and,
consequently, have linearized the dynamics around the attractors,
our approach does not account for the curbing effect of the nonlinear
terms on the amplitude increase.) \bigskip{}

\includegraphics[scale=0.5]{Figure3}

\begin{figure}[H]
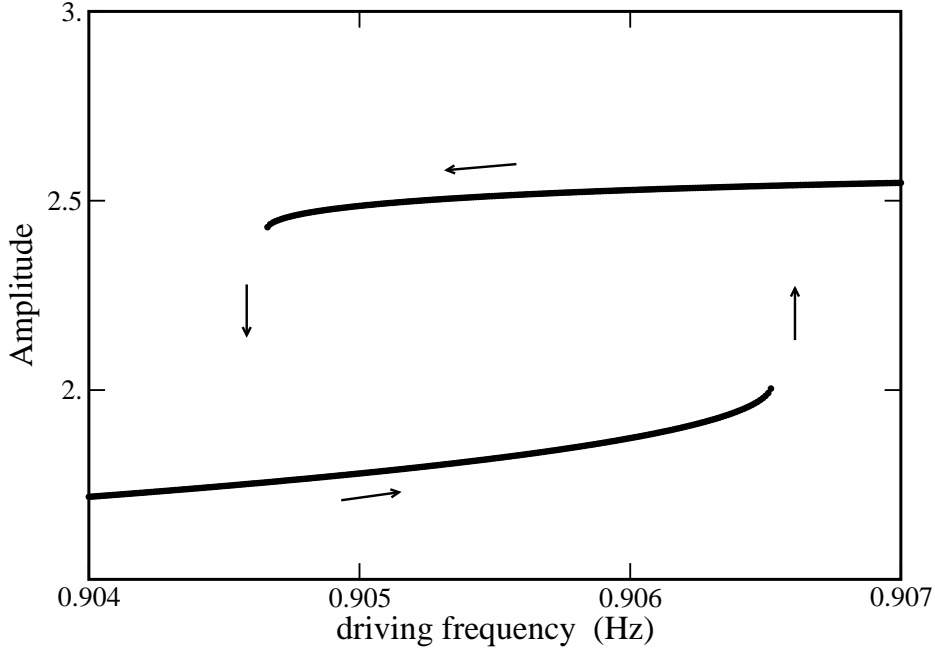

\caption{The amplitude (in arbitrary units) of the system response as a function
of the external frequency. The parameters are the same as those used
in Figs. 1a-1c.}
\end{figure}

Our description also allows us to connect the appearance of prominent
dynamical features with specific locations of the fixed points. It
is worth emphasizing the \emph{singular} character of the points characterized
by $x_{s}=1$ or $x_{s}=1/3$. In them, as can be seen from Eqs. (7),
(8), and (9), one of the equations becomes autonomous, the corresponding
variable entering the other equation as an effective driving. For
those solutions, there is only one (real) characteristic frequency,
which, as shown by Eq. (10), matches the friction coefficient. Hence,
the averaged system, described in terms of $u(t)$ and $v(t)$, is
purely dissipative. Correspondingly, the original (unaveraged) set
{[}$z(t)$, $\dot{z}(t)${]} displays damped oscillations with friction
coefficient $\gamma$ at the frequency $\omega_{d}$. Particularly
relevant to the analysis of the spectral peaks in the noisy system
is the region defined by $1/3\leq x_{s}\leq1$. In that whole range,
(explicitly indicated in Fig. 1), the frequencies $\Omega_{\pm}$
take real values, as can be seen from Eq. (10). Then, the response
of the original system to the driving contains only the frequency
$\omega_{d}$. Moreover, inside that region, an appropriate variation
of $\omega_{d}$ can lead one of the attractors to come closer to
the unstable solution; in that process, the limit $\Delta^{2}(x_{s}-1)(1-3x_{s})-\gamma^{2}\rightarrow0$
is approached, and, eventually, reached at the unstable point. In
contrast, for $x_{s}<1/3$ or $x_{s}>1$, since the frequencies $\Omega_{\pm}$
are complex, the primary-system output incorporates, as components,
damped oscillations at frequencies displaced from $\omega_{d}$. The
partial persistence of this differential behavior in the noisy dynamics
will be linked to the detected emergence of the supernarrow spectral
peaks at $\omega_{d}$: we will see that it is inside the domain of
parameters corresponding to $1/3\leq x_{s}\leq1$ where those peaks
appear.

\section{The response to noise}

Depending on the magnitude of the fluctuations, the stochastic dynamics
can significantly differ from the deterministic picture. For instance,
for the case of bistable behavior, the system is not longer stabilized
in one of the attractors when noise sufficiently intense is added.
Instead, it can switch between the basins, the transition rate being
a function of the noise strength. This is particularly evident in
Fig. 4a, which parallels the noiseless case represented in Figs. 1b
and 2. There, we depict a typical random trajectory which displays
different flips between the two basins. Still, traces of the crucial
role played by the driving frequency in the deterministic dynamics
can be observed in the stochastic evolution. An illustrative example
is presented in Fig. 4b, which corresponds to the parameters of Fig.
1c, i.e., to a frequency value that reduces the distance between the
fixed points $P$ and $R$. There, it is shown that, as $P$ comes
closer to $R$, the basins are deformed, and a preferential direction
for the noise-induced switching becomes patent. In turn, an observable
asymmetric partition develops in the stationary distribution of populations.
In general, the dynamics can be quite complex: the system can display
qualitatively different behaviors depending on the set of parameters.
Here, instead of presenting an exhaustive description of the variety
of possible outputs, we concentrate on gaining insight into the coherent
response to the driving, i.e., into the presence  of distinctively
sharp peaks in the spectrum. To this aim, it is convenient to start
by considering a regime where an analytical description is feasible.
Accordingly, in the following, we will extract some clues to basic
aspects of the dynamics from the study of a linear version of the
model. \bigskip{}

\includegraphics[scale=0.6]{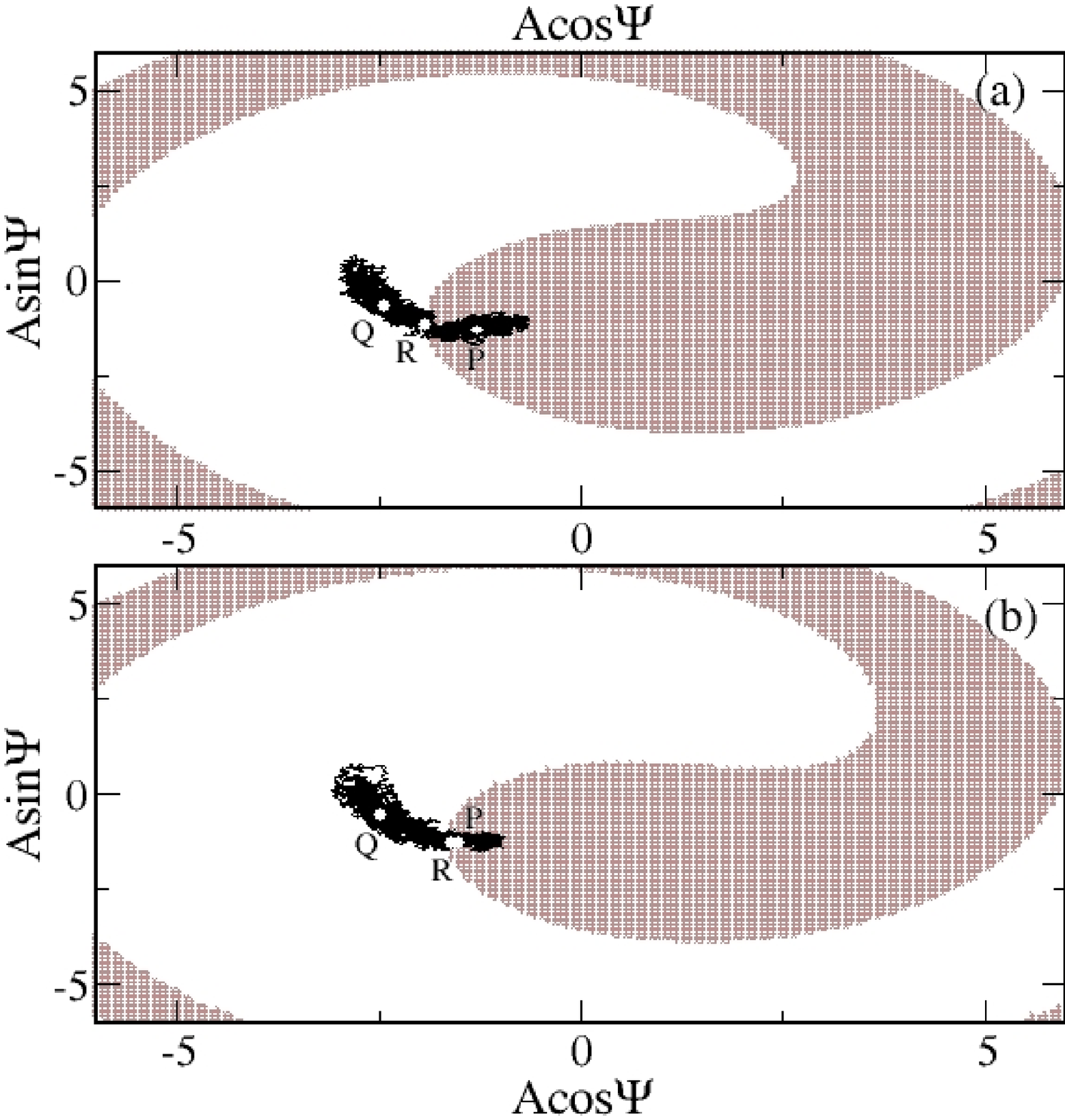}

\begin{figure}[H]
\caption{Noisy trajectories in the effective phase space for the set of parameters
used in Fig. 1b (a); and for the set of parameters used in Fig. 1c
(b). In both cases, the same noise intensity (in arbitrary units)
has been used.}
\end{figure}

\subsection{Analytical approximation to  the spectral density}

The stochastic linearized equations are given by 

\begin{eqnarray}
\dot{u} & = & -\eta_{1}u+\lambda_{1}v+\xi_{1}(t),\\
\dot{v} & = & \lambda_{2}u-\eta_{2}v+\xi_{2}(t),
\end{eqnarray}

\noindent where the diagonal and coupling coefficients take the same
values as in the noiseless case. (The \emph{deterministic} term $\frac{1}{4}\frac{D_{eff}}{\omega_{d}}\frac{1}{A}$,
present in Eq. (3), does not enter the linearized equations since
it has the same order of magnitude as the neglected quadratic terms
in $u$ and $v$. Its role in the dynamics, relevant for increasing
noise intensities, will be discussed later on.) As in the deterministic
version, the cases of having $x_{s}=1$, (or $x_{s}=1/3$), and, in
turn, $\lambda_{1}=0$, (or $\lambda_{2}=0$), correspond to \emph{singular}
behavior: one of the equations represents then an autonomous Ornstein-Uhlenbeck
process \cite{key-RiskenFPequation}, which enters the other equation
as an additive random term. 

To obtain the spectra, we first write the evolution of the position
coordinate of the oscillator using the different changes of variables
and approximations incorporated so far into our description. Namely,
we use the stochastic version of Eq. (12) (now, with noisy $u(t)$
and $v(t)$.) Then, following a standard procedure \cite{key-Stratonovich},
the spectral density $S[z;\omega]\equiv2\int_{-\infty}^{\infty}e^{i\omega\tau}\left\langle z(t)z(t+\tau)\right\rangle d\tau$
is calculated as 

\begin{eqnarray}
S[z;\omega] & = & 2\pi A_{s}\delta(\left|\omega\right|-\omega_{d})+\nonumber \\
 &  & \frac{(\left|\omega\right|-\omega_{d}-\lambda_{1})^{2}+(\left|\omega\right|-\omega_{d}+\lambda_{2})^{2}+2\eta_{1}\eta_{2}}{[\lambda_{1}\lambda_{2}-\eta_{1}\eta_{2}+(\left|\omega\right|-\omega_{d})^{2}]^{2}+4(\left|\omega\right|-\omega_{d})^{2}\eta_{1}\eta_{2}}\frac{D_{eff}}{2}
\end{eqnarray}
The $\delta$-component corresponds to the term $A_{s}\cos(\omega_{d}t+\Psi_{s})$
in Eq. (12), i.e., it represents a purely deterministic output at
$\omega_{d}$; its weight in the spectrum is determined by the amplitude
at the corresponding attractor. The second component of $S[z;\omega]$
incorporates the effect of the fluctuations, present in Eq. (12) through
$u(t)$ and $v(t)$. We will focus on this second term, since it was
in the (continuous) noisy component of the spectrum where the supernarrow
peaks were observed. (The $\delta$-component was removed in the procedure
followed to analyze the experimental spectra \cite{key-Stambaugh}.)
\bigskip{}

\includegraphics[scale=0.6]{Figure5}

\begin{figure}[H]
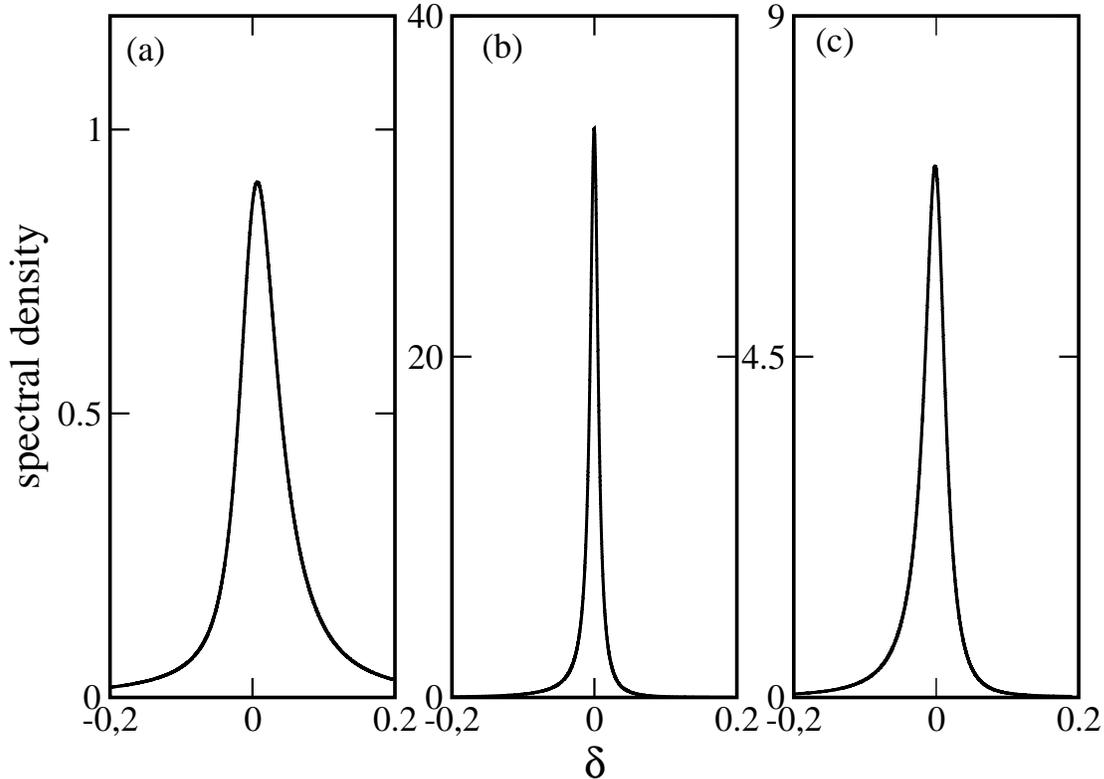

\caption{Spectral density (in arbitrary units) as a function of $\delta=\omega-\omega_{d}$
(in $Hz$) for three different values of the driving frequency: $\omega_{d}=0.902\; Hz$
(a); $\omega_{d}=0.9056\; Hz$ (b); and,  $\omega_{d}=0.909\; Hz$
(c).  The rest of parameters are the same as those used in Figs. 1a-1c.}
\end{figure}

Results for the spectral density obtained from Eq. (18) for three
different values of $\omega_{d}$ are represented in Fig. 5. (Note
that different vertical scales were used to obtain the graphics.)
The connection between the spectral features and the characteristics
of the dynamics is established in the following points: 

i) As found in the experiments, the spectrum is not symmetric with
respect to $\omega_{d}$. This is apparent in the lack of symmetry
of the numerator in Eq. (18), and it is a consequence of the correlation
existent between the fluctuations in the amplitude and in the phase,
{[}see Eq. (4).{]} It is also noticeable in Eq. (18) that, for generic
system parameters, the second term in the spectrum does not necessarily
peak at $\omega_{d}$; in fact, more than one peak can be present.
The localization and width of those additional peaks change as $\omega_{d}$
is varied. These results, which account for the presence of secondary
maxima in the observed spectra, can be understood from our analysis
of the noiseless dynamics. It was found that, in a general regime,
the deterministic variables $u(t)$ and $v(t)$ can present complex
characteristic frequencies $\Omega_{\pm}$, whose imaginary components
imply the occurrence of shifts in the frequency of the complete-system
output. Now, we can identify those shifts as leading to spectral peaks
displaced from $\omega_{d}$. 

ii) Useful for the characterization of the peaks at $\omega_{d}$
is the introduction of the effective width $\Gamma$, defined from
Eq. (18) by the expression

\begin{eqnarray}
\Gamma & = & \sqrt{\left|\lambda_{1}\lambda_{2}-\eta_{1}\eta_{2}\right|},
\end{eqnarray}
which is cast into 

\begin{equation}
\Gamma=\sqrt{\left|\Delta^{2}(x_{s}-1)(1-3x_{s})-\gamma^{2}\right|}
\end{equation}
for our particular system. The coherent response to the driving force
is marked by the occurrence of a small  value of $\Gamma$. {[}Since
the different parameters that enter Eq. (20) take finite values, it
is the combination of them leading to a small $\Gamma$ that induces
the appearance of the supernarrow peaks.{]} From the form of $\Gamma$,
closely related to Eq. (10), the relevance of the former analysis
of the stability to the current discussion is evident. Indeed, the
mechanism responsible for the sharpening of the peaks can be traced
from previous arguments. As the frequency is varied, the localization
of the fixed points changes. In particular, the region $1/3\leq x_{s}\leq1$
can be entered. In that domain, there is a preferential response of
the system at the driving frequency. Moreover, there is a significant
reduction in $\Gamma$. The limit case of having $\Gamma=0$, i.e.,
of working with parameters that fulfill the equation $\Delta^{2}(x_{s}-1)(1-3x_{s})-\gamma^{2}=0$,
corresponds to the convergence of one of the attractors on the unstable
point. Then, the supernarrow peaks are rooted in the peculiar behavior
of the noiseless system when the unstable point inside the region
$1/3\leq x_{s}\leq1$ is approached. (Here, a comment on the applicability
of our description is pertinent. Given that the linearization of the
stochastic equations applies to regimes of significant localization
around the attractors, and, therefore, to weak noise, the description
becomes less valid as the unstable point is approached. Even so, a
sound picture of general trends connected to the observed effects
is outlined in our framework.) 

It is important to take into account that a small increment in $\omega_{d}$
can qualitatively alter the dynamics. For instance, by slightly shifting
the frequency, the whole region $1/3\leq x_{s}\leq1$ can be traversed,
and, in turn, significant changes in localization and stability can
be induced. (This is apparent in Fig. 1: in the considered region
of intersection between the graphics, small changes in the ordinate
at the origin and in the slope of $g(x)$ can considerably alter the
set of fixed points.) Then, one can understand the  abrupt appearing
of the supernarrow peaks, detected through the frequency variation,
and qualitatively reproduced in Fig. 5. It follows that the practical
realization of the supernarrow-peak regime can require considerable
precision in fixing $\omega_{d}$. 

Additional support to the above arguments is provided by the analysis
of the peak height $Q_{\omega_{d}}$ i.e., the value of the noisy
component of $S[z;\omega]$ at $\omega_{d}$, which is given by 

\begin{equation}
Q_{\omega_{d}}=\frac{\lambda_{1}^{2}+\lambda_{2}^{2}+2\gamma^{2}}{\Gamma^{4}}\,\frac{D_{eff}}{2}.
\end{equation}
In Figs. 6a and 6b, we respectively depict $\Gamma$ and $Q_{\omega_{d}}$
as a function of $\omega_{d}$. Following the same sequence as in
Fig. 1, we start with a regime where there is only one attractor,
$P$. As the frequency is increased, $\Gamma$ diminishes, and, in
turn, the peak height grows. Eventually, the second attractor $Q$
turns up. In the central region where $P$ and $Q$ coexist, a substantial
sharpening of the peaks can be observed. Although it is not equal
in both attractors, the width reduction is significant in any of them.
As $\omega_{d}$ grows further, only the point $Q$ survives, and
the associated $\Gamma$ displays a monotonous growth in the depicted
domain of frequencies. The singularities in $Q_{\omega_{d}}$ mark
the appearance (or disappearance) of one of the attractors via its
convergence on the unstable point. It is in the specific range of
frequencies between the two singularities where the coherent response
is strongly localized. The form of the graphics in the central domain
points to a coherent response at distances from the unstable point
sufficiently large to guarantee the stability at weak noise: the widths
(heights) of the peaks are small (large) enough to imply coherence
in the dynamics. {[}The crossing points of the graphics provide illustrative
examples of this situation: they are distant from the unstable solution
and yet present significantly small (large) values of $\Gamma$ ($Q_{\omega_{d}}$).{]}
Hence, despite the noise-induced flips between basins, the system
can be expected to spend considerable time in regions, near $P$ or
$Q$, where the behavior is coherent.

The presence of approximately equal populations in the attractors,
identified  in a preliminary analysis of the experimental results
as an indicator of the supernarrow-peak regime, can be explained within
our approach. In the regime of small $\Gamma$, i.e., inside the region
$1/3\leq x_{s}\leq1$, the basins have similar sizes, and, consequently,
any uniform initial arrangement of population ends up as an approximately
symmetric partition between the two attracting sites of the system.
Departures from symmetry take place as the borders of the central
region in Fig. 6 are approached. 

\bigskip{}
\includegraphics[scale=0.6]{Figure6}

\begin{figure}[H]
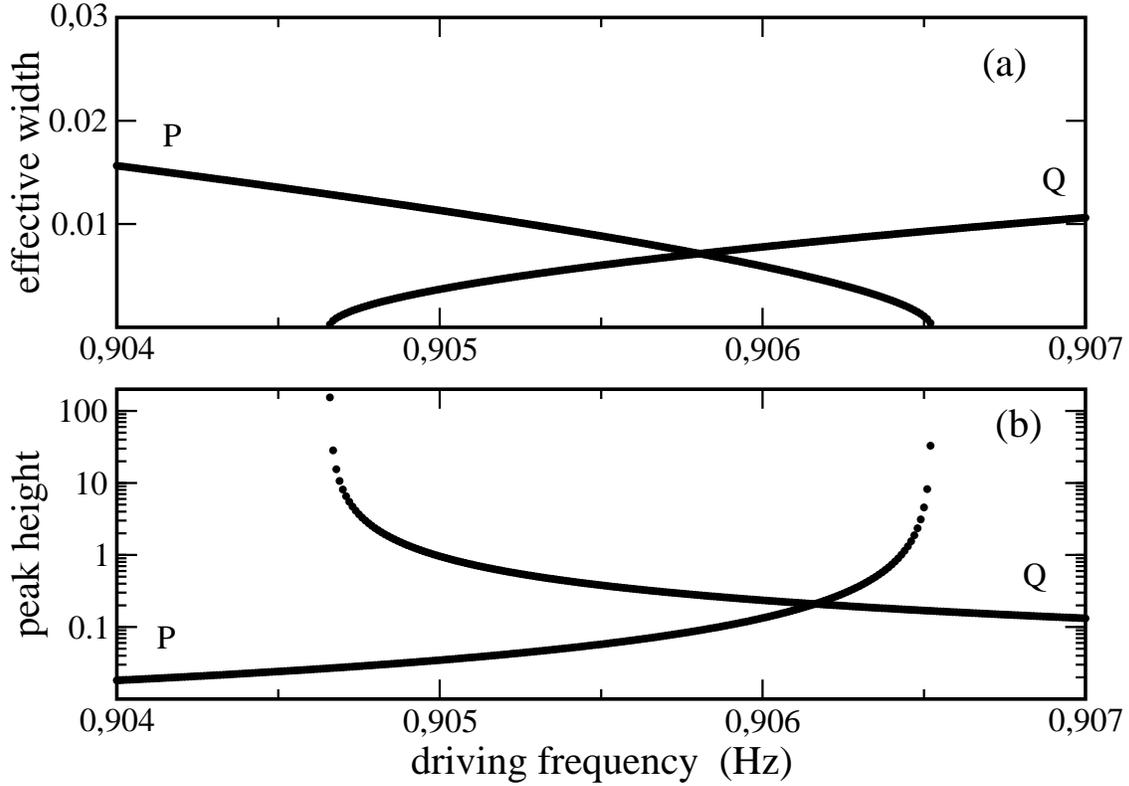

\caption{Effective width (a) and height of the spectral peak (b) (both in arbitrary
units) as a function of the driving frequency. The parameters are
the same as those used in Figs. 1a-1c.}
\end{figure}

iii) Special care is needed in assessing the persistence of the supernarrow
spectral peaks as the magnitude of the fluctuations is enhanced. For
a growing noise intensity, the rate of switching between the basins
increases, and the dependence of the spectral density on the effective
noise strength $D_{eff}$ becomes nontrivial. (Recall that the spectral
density given by Eq. (18) is strictly valid in the considered weak-noise
limit.) For stronger fluctuations, apart from an increasing number
of noise-induced transitions, the system develops a less coherent
behavior inside each basin. As a result, the peaks broaden. Some insight
into this behavior is obtained from the analysis of the role of the
noise-induced \emph{deterministic} term $\frac{1}{4}\frac{D_{eff}}{\omega_{d}}\frac{1}{A}$
present in Eq. (3) and of the strictly random component of Eq. (4),
$\frac{\xi_{2}(t)}{A}$. Both terms have a stronger effect as the
localization of the fixed points approaches smaller amplitudes. (An
implication related to this noise-induced \emph{asymmetry} is that
the \emph{noisy} trajectories do not reach the point with $A=0$.)
The term $\frac{1}{4}\frac{D_{eff}}{\omega_{d}}\frac{1}{A}$ modifies
the position of the fixed points obtained from the purely deterministic
picture. From a scenario of extended applicability of the linearized
equations, this term can be reasonably conjectured to change also
the parameters that determine the effective width of the peaks in
the spectrum, and, consequently, to prevent reaching small values
of $\Gamma$. Additionally, the multiplicative character of the random
term can significantly affect the form of the spectral density. Then,
although our approach does not describe the transitions between attracting
sites, it is primarily understood from it that, as the noise intensity
grows, the coherence in the response to the driving decays. Still,
given the robust character of their deterministic roots, i.e., of
the peculiar  dynamics near each of the attractors, the peaks can
be expected to be observable for moderate noise intensities.

\subsection{General aspects of the supernarrow-peak regime}

We turn now to evaluate the generality of the considered effects.
From the above analysis, a procedure to identify similar features
in other driven systems can be established. Moreover, the criteria
for their emergence can be defined. We consider a \emph{generic} system,
where, as preliminary requirements, it is assumed that a coarse-grained
reduction is feasible and that a linearization of the resulting effective
stochastic equations applies. Hence, we deal with a system described
by Eqs. (16) and (17) with general coefficients $\eta_{i}$ and $\lambda_{i}$,
($i=1,2$). From the following arguments, sufficient conditions for
the appearance of the supernarrow peaks are set up: 

First, given that the linearization method implies a restriction of
the dynamics to the areas around the fixed points, the stability requirements,
given by Eqs. (13) and (14), must hold. Second, since a preferential
output of the system at the driving frequency is intended, we must
look for characteristic frequencies of the averaged system which take
real values. (This corresponds to have only the frequency $\omega_{d}$
in the original-system response.) From Eq. (10), it is apparent that
the frequencies $\Omega_{\pm}$ are real when the inequality $(\eta_{1}-\eta_{2})^{2}+4\lambda_{1}\lambda_{2}>0$
holds. A final requirement comes from imposing a reduced width for
the spectral peaks. Namely, a small value of $\Gamma=\sqrt{\left|\lambda_{1}\lambda_{2}-\eta_{1}\eta_{2}\right|}$
is needed to guarantee the sharpening of the spectral peaks. 

This procedure implies the control of the elements of the system which
are essential for the appearance of a coherent response to the driving.
The resulting set of restrictions defines the aimed range of parameters.
(Corrections to the simplified picture given by our approach must
be implemented when the system presents a highly-nonlinear term or
incorporates strong fluctuations.) 

It is worth mentioning that an additional element of generality is
provided by the compact characterization of the nonlinear oscillator
specifically considered in our study: the coefficients $c_{1}$ and
$c_{2}$ embody all the influence of the system elements on the analyzed
effects. This compact character facilitates setting up a common framework
for the analysis of different systems. (Examples can be found in \cite{key-Stratonovich}.)

\section{concluding remarks}

Our approximate description of the stochastic dynamics of a damped
and driven nonlinear oscillator has allowed us to establish a link
between the appearance of distinctively supernarrow peaks in the spectrum
and specific aspects of the deterministic dynamics. Although the preferential
response  at the driving frequency, marked by the appearance of supernarrow
spectral peaks, occurs in the domain of bistability, it is the dynamics
near each attracting site that accounts for the observed coherence.
The peaks have been shown to become more prominent when, with the
system still inside the stability domain, one the attractors approaches
the unstable solution. One significant implication of these findings
is that, for those features to persist when the noise intensity is
increased, significant permanence in the stability region must be
guaranteed. Actually, the coherent response seems to be robust against
the dephasing effects of the noise-induced transitions between basins:
the deterministic roots of the effect are still patent for growing
noise strength. 

We stress that despite the use of scaled parameters (convenient for
the calculations) and of different approximations, the analysis reproduces
the main features detected in the experiments. Namely, the hysteresis
curve for the amplitude versus the frequency, the form of the spectral
densities, and the similar values of the populations in the attractors
in the supernarrow regime are satisfactorily simulated. 

The supernarrow spectral peaks have been reproduced with a model where
the noise correlation-time has been considered to be much smaller
than any other time scale in the system. In fact, this is a requirement
for the validity of the stochastic version of the applied averaging
method. Hence, there are no qualitative differences between the broadband
regime, considered in our approach, (and realized in the experiments),
and the strict white-noise limit. The analysis of the role of specific
colored-noise characteristics requires going beyond the used coarse-grained
approach.

The study uncovers the generality of the mechanism behind the observed
effects, and, therefore, its relevance to different contexts. The
compact character of our description facilitates the possibility of
controlling the setup, in particular, of varying the components to
systematically characterize the response under well-defined conditions.
The realization of similar effects in unexplored regimes is then open.

\end{document}